# PREDICTORS OF JAVA PROGRAMMING SELF–EFFICACY AMONG ENGINEERING STUDENTS IN A NIGERIAN UNIVERSITY

By

Philip Olu Jegede, PhD
Institute of Education
Obafemi Awolowo University, Ile-Ife, Nigeria

## ABSTRACT

*The study examined the relationship between Java programming self-efficacy and programming background of engineering students in a Nigerian University. One hundred and ninety two final year engineering students randomly selected from six engineering departments of the university participated in the study. Two research instruments: Programming Background Questionnaire and Java Programming Self-Efficacy Scale* were *used in collecting relevant information from the subjects. The resulting data were analyzed using Pearson product correlation and Multiple regression analysis. Findings revealed that Java Programming self-efficacy has no significant relationship with each of the computing and programming background factors. It was additionally obtained that the number of programming courses offered and programming courses weighed scores were the only predictors of Java self-efficacy.*

## INTRODUCTION

In a recent study, Askar and Davenport [1] identified variables that are related to self-efficacy of engineering students in Turkey, concluding with factors such as gender, computer experience, and family usage of computers. The importance of the study was based on the necessity of computer skills for today's engineering professional practices and the factors that would affect their ability to acquire programming skills. However literatures and classroom experience have suggested other factors that may be associated or impact upon programming self-efficacy. For example Romalingans, La Belle and Wiedenbeck [2] posited that programming self-efficacy is often influenced by previous programming experience as well as mental modeling.

Bandura [3] posited that judgments of self-efficacy are based on four sources of information. The sources included individual performance attainments, experiences of observing the performance of others, experiences of observing the performance of others, verbal persuasion and psychological reactions that people use partly to judge their capability. This is also applicable to programming domain. Performance attainment in this context can be measured by the scores of students in programming courses. In other words if students had persistently scored reasonably in previous programming courses, they tend to increase in their self efficacy. If research can identify predicting factors of programming self-efficacy, the problem of poor performance in programming as well as that of approach avoidance of programming in the future professional practice can be solved particularly among engineers of today as they are daily confronted with tasks that are computer and software driven. Studies identifying discrete factors that are related to programming self efficacy are lacking in Nigeria. Identifying success criteria for computer programmers can help improve training and development programs in academic and industrial settings [4]. However no study can investigate the self-efficacy of all programming languages at a time. Thus this study starts with Java programming language as one of the object oriented languages recently introduced into the curricula of some engineering departments in Nigeria. Other object-oriented programming languages replacing the procedural ones in the old curricula include Matlab, $C^{++}$ and $C^{\#}$. The goal of this work therefore is to study Java self-efficacy of engineering students by exploring the relationship between Java self-efficacy and each of computing background, programming experience in years and programming courses weighed scores, number of programming courses taken. The study also seeks to investigate their combined influence on Java self-efficacy. Specifically; the study will answer the following questions.

1. What is the relationship between Java self-efficacy and each of computing background, Programming experience, programming weighed scores and number of programming courses taken?
2. Will a combination of these selected factors significantly predict Java self-efficacy?
3. What is the proportion of variance in Java self-efficacy accounted for by the linear combination of the factors; computing experience, programming experience, programming weighed score and number of programming courses taken?
4. What is the relative contribution of each factor in the prediction of Java self-efficacy?

## METHOD

One hundred and ninety two final year students who offered programming randomly selected from six engineering departments of Obafemi Awolowo University, Ile-Ife, Nigeria participated in the study. These included Mechanical, Civil, Metallurgy and Material Engineering departments; others include Electrical, Chemical and Computer Science and Engineering departments.

Two research instruments were employed to collect relevant data from the students. These were Programming Background Questionnaire (PBQ) and Java Programming Self-efficacy Scale (JPSES). PBQ was designed to obtain information on engineering students programming experience, number of programming courses previously undergone and scores obtained in those programming courses. JPSES was developed from the computer programming self-efficacy scale of Ramalingam and Wiedenbeck [2] by Askar and Davenport [1]

Participants were to rate their confidence in performing some specified Java programming related tasks. The confidence was to be rated for each item in a seven –point Likert scale as following:  Not confident at all (1), Mostly not confident (2), Slightly confident (3), Averagely confident (4), Fairly confident (5), Mostly confident (6), Absolutely confident (7). Total score obtainable on the said efficacy scale was 224 while the minimum score totaled 32.The instruments were administered on the students with the assistance of their lecturers. The resulting data were analyzed using Pearson product correlation and Multiple regression Analysis.

# RESULTS

**Table 1: Relationship between Java self-efficacy and Computing and Programming Background**

|  | Computing Experience | Year of First Programming | Weighed Score in Programming Courses | Number of Programming Courses taken |
|---|---|---|---|---|
| Java Self-Efficacy | -.029 | .099 | .278 | .453 |

From Table 1, the correlated coefficient between Java programming self-efficacy and each of computing experience, year of first programming, weighed scores in programming courses and number of programming courses taken were each found to be r= -.029, .099, .278 and .453. The relationship was not significant at .05 level of significance.

**Table 2: Summary of Analysis of Variance of Programming Background and Java Programming Self-Efficacy**

**ANOVA**[b]

| Source of Variance | Sum of Squares | Df | Mean Square | F | Sig. |
|---|---|---|---|---|---|
| 1  Regression | 148157.887 | 4 | 37039.472 | 19.821 | .000[a] |
| Residual | 351306.828 | 188 | 1868.653 |  |  |
| Total | 499464.715 | 192 |  |  |  |

**Table 3: Summary of Multiple Regression Analysis of the Relationship between Java Programming Self-Efficacy and Programming Background**

| Variables Entered | R | R Square | Adjusted R Square | Std. Error of the Estimate | Sig. |
|---|---|---|---|---|---|
| Experience in computing<br>Year of first programming<br>Nunmber of program<br>Average score | .545[a] | .297 | .282 | 43.22792 | .000 |

**Table 4: Significant tests of Regression Weights of Independent Variables**

**Coefficients[a]**

| Model | | Unstandardized Coefficients | | Standardized Coefficients | T | Sig. |
|---|---|---|---|---|---|---|
| | | B | Std. Error | Beta | | |
| 1 | (Constant) | -80.003 | 24.752 | | -3.232 | .001 |
| | Experience in computing | -4.758 | 3.530 | -.085 | -1.348 | .179 |
| | year of first programming | 1.950 | 2.568 | .047 | .759 | .449 |
| | Number of program | 26.548 | 3.530 | .469 | 7.520 | .000 |
| | Average score | 1.482 | .337 | .272 | 4.397 | .000 |

a. Dependent Variable: Java self efficacy

To verify whether a combination of the computing and programming related background variables will significantly predict Java self-efficacy, data obtained from programming background questionnaires and Java self-efficacy scale were subjected to multiple regression analysis. Table 2 shows the summary of the analysis of variance of the independent variables in the regression procedures.

The results in Table 2 show that the analysis of variance of the multiple regression data yielded an F-ratio of 19.821 which is significant at .05 level. This implies that a combination of the independent variables (ie. Computing experience, programming experience in years, number of programming courses taken and the average score in the programming courses) is significantly related to Java self-efficacy of the engineering students.

The results of the regression analysis on the relationship between the dependent variable and the combination of the four independent variables are as stated in Table 3, the table shows that using the four independent variables (computing experience, year of first programming number of programming courses and the average score in programming courses) to predict Java programming self-efficacy gives a coefficient of multiple regression ® of .545 and a multiple correlation square ($R^2$) of .297. These values are statistically significant at .05 level, which suggests that only 29.7 percent of the variance of Java self-efficacy were explained by the by the combination of the four independent variables. Further attempt was made to determine the relative power of each of the independent variables to predict Java self-efficacy of engineering students. Table 4 shows, for each of the variables, Error of Estimate (SEB), Beta, T-ratio and the level at which T-ratio is significant.

From the table, the number of programming courses taken and the average score in programming courses taken had t-values of 7.520 and 4.397 respectively. The values of Beta-weights for the two variables are .469 and .272 respectively. These values are significant at .05 level of confidence which implies that the two variables contribute majorly to the prediction of Java self-efficacy. From the values of Beta weights and t-ratios for each independent variable, it is clear that the number of programming courses offered had the highest impact in the prediction of Java programming self-efficacy followed by the average score of the programming courses offered. Year of first programming and experience in computing had t-values and Beta weights that are not significant at the .05 level. Summarily, it could be said that the number of programming courses taken and the average score of programming courses offered by engineering students had significant contributions to the prediction of Java self-efficacy, the weights of experience in computing and year of first programming demonstrated week contribution

## DISCUSSION

The study founds that the number of programming courses offered by students and their achievements in the programming courses (based on scores) significantly predict their Java programming self-efficacy. This appear consistent with the position of Wiedenbeck [5] who obtained that previous programming experience affected perceived self-efficacy on one hand and that perceived self-efficacy in programming also affected performance in programming courses.

In an earlier study Ramalingan, La Belle and Wiedenbeck [2] had come out with the results that self-efficacy for programming were influenced by previous programming experience. Bandura [3] also opined that self-efficacy perceptions develop gradually with the attainment of skills and experience. The fact that self-efficacy in programming domain becomes predictable by performance in programming course is logical. This is because learners with high self-efficacy are more likely to undertake challenging tasks and to expend considerably greater efforts to complete them in the face of unexpected difficulties, than those with lower self-efficacy [1]

However, the number of years a student had been introduced to programming did not significantly predict Java self-efficacy. This can be understood in this way; experiences in programming by years may not necessary imply continuous active programming experience for example many of the engineering students in the various departments used for the study did offer for the first time programming courses in their second year. The secondary school curriculum in Nigeria do not accommodate programming content and it would be quite unlikely that students took initiative to learn programming on their own before gaining admission into the university. Thus the subjects used for the study appear to experience programming approximately around the same time. Apart from this, students might not get involved in programming except in the semester during which programming as a course was compulsory, hence years of programming experience did not predict Java self-efficacy.Similarly, years of computing experience did not predict Java self-efficacy, this is perhaps because the substantial part of the skills acquired in the course of students encounter with computers used for the study were not in programming domain. Rather many of these skills were internet and word processing-related. This opposed the findings of Askar &Davenport [1] who posited that the number of years of experience a student had with computers had a significant linear contribution to their self-efficacy scores.

The findings above have pedagogical implications. Educational researchers recognize that because skills and self-efficacy are so interwined,one way of improving student performance is to improve student self efficacy [6]. Wiendebeck,et al,[6] believed that students must steadily carry out tasks of increasing difficulty, until they have a history of solid attainments. Expounding more on this idea, increasing performance through self efficacy in programming courses will necessitate the following;

(i)  More assignments at the beginning of the programming courses than at the end of the semester. The assignment should move gradually from simple to complex given severally. Observation has shown that instructors often wait till the end of the semester (i.e. close to the examination) before giving students assignments. But when assignments are given severally at the beginning of the course, confidence of students become boosted particularly when the assignments are undertaken with success.

(ii) Prompt feedbacks must be ensured; even when students undertake regular assignment and their scores are not made known promptly, reason(s) for the assignments become defeated, on the other hand performance accomplishment becomes assured when students receive prompt feedback with success scores thereby leading to higher self-efficacy.

(iii) In the course of instructional lessons, group work in programming classes would help increase self efficacy. This is because experiences of observing the performance of others give rise to self efficacy. This is as posited by Bandura [3].

**CONCLUSION**

This study obtained that weighed scores in programming courses and the number of programming courses offered by engineering students were the significant predictors in Java programming self-efficacy. This study also finds no significant relationship between Java programming self-efficacy and each of engineering students computing background and years of first programming. Further studies are needed in identifying factors that will better predict Java self efficacy. In addition to this, the study need be replicated for other object oriented languages currently introduced into the curriculum. A possible limitation of the study was that the scores obtained in programming courses did not derive from standardized tests. They were proceeds of teacher-made tests with their inherent weaknesses.

## ACKNOWLEDGEMENT

The study acknowledged Askar & Davenport [1] whose works provided an instrument and inspiration for this effort.

## AUTHOR'S PROFILE

Dr Philip Jegede is an Associate Professor in the Institute of Education of Obafemi Awolowo University,Ile-Ife,Nigeria.He holds both Bachelor and Master of Science degrees in Mathematics from University of Lagos,Nigeria.He later ventured into the field of Education by enrolling and completing a Master of Education and consequently a PhD degree in Curriculum Studies with focus in ICT. His research interest is in Computer Education. Before his present appointment, he had lectured in a College of Education and a Polytechnic School.